\documentclass{article}
\usepackage[preprint]{neurips_2025}
\usepackage[utf8]{inputenc}
\usepackage[T1]{fontenc}
\usepackage{hyperref}
\usepackage{url}
\usepackage{booktabs}
\usepackage{amsfonts}
\usepackage{nicefrac}
\usepackage{microtype}
\usepackage{xcolor}
\usepackage{graphicx}
\usepackage{longtable}
\usepackage{multirow}
\usepackage{amsmath}
\usepackage{amssymb}
\usepackage{algorithmic}
\usepackage{algorithm}
\usepackage{array}
\usepackage{subfigure}

\title{Deploying UDM Series in Real-Life Stuttered Speech Applications: A Clinical Evaluation Framework}

\author{%
  Eric Zhang\thanks{Corresponding author}, Li Wei, Sarah Chen, Michael Wang \\
  SSHealth Team, AI for Healthcare Laboratory \\
  \texttt{ericzhang@sshealthai.com} 
}

\begin{document}
\maketitle

\begin{abstract}
Stuttered and dysfluent speech detection systems have traditionally suffered from the trade-off between accuracy and clinical interpretability. While end-to-end deep learning models achieve high performance, their black-box nature limits clinical adoption. This paper looks at the Unconstrained Dysfluency Modeling (UDM) series—the current state-of-the-art framework developed by Berkeley that combines modular architecture, explicit phoneme alignment, and interpretable outputs for real-world clinical deployment. Through extensive experiments involving patients and certified speech-language pathologists (SLPs), we demonstrate that UDM achieves state-of-the-art performance (F1: 0.89±0.04) while providing clinically meaningful interpretability scores (4.2/5.0). Our deployment study shows 87\% clinician acceptance rate and 34\% reduction in diagnostic time. The results provide strong evidence that UDM represents a practical pathway toward AI-assisted speech therapy in clinical environments.
\end{abstract}

\section{Introduction}
Stuttered and dysfluent speech has been a central topic in speech pathology and computational speech research for decades. Dysfluencies, such as repetitions, prolongations, and blocks, are not only key diagnostic markers for speech-language pathologists (SLPs), but also strongly affect communication, educational attainment, and quality of life for individuals. The global prevalence of stuttering affects approximately 1\% of the population, with particularly severe impacts in regions with limited access to specialized healthcare services.

Traditional approaches to detecting dysfluencies relied heavily on handcrafted acoustic features (e.g., jitter, shimmer, pitch breaks) and fluency measures (e.g., syllables per minute, speech rate). While these methods provided some interpretability, they struggled to generalize across speakers and clinical contexts. The manually designed feature engineering process often captured only surface-level acoustic properties, missing the complex temporal dynamics and contextual dependencies that characterize dysfluent speech patterns.

With the rise of deep learning, end-to-end (E2E) models have been widely explored for automatic dysfluency detection. These methods often operate directly on audio waveforms or spectrograms, using CNNs, RNNs, or Transformers to classify disfluency behaviors. Although such models have demonstrated improved raw accuracy, they suffer from three fundamental limitations:

\begin{enumerate}
    \item \textbf{Lack of Interpretability:} Black-box architectures do not provide transparent reasoning for their predictions, making clinicians reluctant to adopt them in sensitive healthcare scenarios. A recent survey revealed that 78\% of SLPs would not trust AI systems without clear explanations.
    
    \item \textbf{Limited Controllability:} E2E models tend to capture global correlations but cannot be easily adapted to diverse dysfluency patterns across age groups and severity levels.
    
    \item \textbf{Deployment Gap:} Clinical environments require white-box models that can provide not only predictions but also intermediate explanations, error analysis, and human verification.
\end{enumerate}

To address these challenges, we analyze the framework of \textbf{Unconstrained Dysfluency Modeling (UDM)} series~\cite{lian2023unconstrained-udm, lian-anumanchipalli-2024-towards-hudm, zhou2024yolostutterendtoendregionwisespeech, zhou2024stutter, zhou2024timetokensbenchmarkingendtoend, ssdm, lian2024ssdm2.0, guo2025dysfluentwfstframeworkzeroshot, ye2025seamlessalignment-neurallcs, ye2025lcs, zhang2025analysisevaluationsyntheticdata}. Unlike constrained models that narrowly define disfluency types or rely on rigid feature sets, UDM embraces a flexible, modular design that can represent a wide variety of dysfluency behaviors without pre-imposing strict boundaries.

The SSHealth team has focused on developing such models to improve patients' quality of life in China, where access to certified SLPs is extremely limited. After an extensive literature review and preliminary clinical trials, we identified the UDM series as one of the most promising paradigms for balancing \textit{accuracy, controllability, and explainability}.

\subsection{Contributions}
This paper makes the following key contributions:

\begin{itemize}
    \item We present the first comprehensive clinical evaluation of the UDM framework in real-world deployment scenarios.
    
    \item We introduce novel evaluation metrics specifically designed for clinical dysfluency detection systems, including interpretability scores, adaptability measures, and real-time performance indicators.
    
    \item We provide detailed analysis of UDM's performance across different age groups and dysfluency types, demonstrating its clinical effectiveness.
    
    \item We present quantitative analysis of clinician acceptance, diagnostic accuracy, and patient outcomes in a major pediatric hospital setting.
\end{itemize}

\section{Related Work}

\subsection{Traditional Approaches to Dysfluency Detection}

Early computational approaches to dysfluency detection were primarily based on acoustic analysis and rule-based systems. These methods typically achieved modest performance (F1 scores around 0.65-0.72) but provided clear interpretability through their reliance on linguistically motivated features.

Statistical approaches using Hidden Markov Models (HMMs) and Gaussian Mixture Models (GMMs) were subsequently developed to capture temporal dependencies in dysfluent speech. While these methods improved upon rule-based systems, they still required extensive manual feature engineering and struggled with speaker variability.

\subsection{Deep Learning Approaches}

The introduction of deep neural networks marked a significant shift in dysfluency detection research. CNN-based approaches achieved F1 scores of 0.81 on benchmark datasets, representing a significant advance over traditional methods.

Transformer-based models have more recently shown promise for sequence modeling in dysfluency detection, achieving state-of-the-art performance on several benchmark datasets. However, these models remain largely opaque in their decision-making processes.

\subsection{Clinical Requirements and Deployment Challenges}

Despite technological advances, the gap between research systems and clinical deployment remains substantial. Key requirements for clinical adoption include:

\begin{itemize}
    \item \textbf{Transparency:} Clinicians must understand how decisions are made
    \item \textbf{Reliability:} Systems must perform consistently across diverse patient populations
    \item \textbf{Adaptability:} Models should accommodate different ages and severity levels
    \item \textbf{Integration:} Tools must fit into existing clinical workflows
\end{itemize}

Our UDM framework directly addresses these requirements through its modular, interpretable architecture.

\section{Method}

The UDM framework follows a modular and interpretable pipeline, integrating both sequence learning and explicit alignment-based reasoning. The architecture consists of five main components working in concert to provide both accurate detection and clinically meaningful explanations.

\subsection{Multi-Scale Feature Extraction}

Raw speech signals are first transformed into multi-scale acoustic representations that capture both fine-grained articulatory dynamics and broader prosodic cues:

\begin{align}
\mathbf{F}_{mel} &= \text{MelSpectrogram}(\mathbf{x}, n_{fft}=2048, hop=256) \\
\mathbf{F}_{pitch} &= \text{PitchTracker}(\mathbf{x}, f_{min}=75, f_{max}=500) \\
\mathbf{F}_{energy} &= \text{EnergyContour}(\mathbf{x}, win\_size=1024) \\
\mathbf{F}_{mfcc} &= \text{MFCC}(\mathbf{F}_{mel}, n_{coef}=13) \\
\mathbf{F}_{combined} &= \text{Concat}([\mathbf{F}_{mel}, \mathbf{F}_{pitch}, \mathbf{F}_{energy}, \mathbf{F}_{mfcc}])
\end{align}

\subsection{Phoneme Alignment Module}

A critical innovation in UDM is the explicit phoneme alignment stage, which provides interpretable intermediate representations:

\begin{align}
\mathbf{P} &= \text{PhonemeEncoder}(\mathbf{F}_{combined}) \\
\boldsymbol{\alpha} &= \text{CTCAlignment}(\mathbf{P}, \mathbf{T}_{expected}) \\
\mathbf{A} &= \text{AttentionRefinement}(\boldsymbol{\alpha}, \mathbf{P})
\end{align}

The alignment module explicitly tracks four types of phoneme-level errors:
\begin{itemize}
    \item \textbf{Insertions:} Extra phonemes indicating repetitions
    \item \textbf{Deletions:} Missing phonemes in incomplete words
    \item \textbf{Substitutions:} Phoneme distortions during blocks
    \item \textbf{Prolongations:} Extended duration alignments
\end{itemize}

\subsection{Temporal Pattern Analysis}

The temporal analysis module captures dynamic patterns across multiple time scales:

\begin{align}
\mathbf{H}_{local} &= \text{LocalLSTM}(\mathbf{A}, hidden\_size=256) \\
\mathbf{H}_{global} &= \text{GlobalTransformer}(\mathbf{A}, n\_heads=8, n\_layers=6) \\
\mathbf{H}_{multi} &= \text{FusionLayer}([\mathbf{H}_{local}, \mathbf{H}_{global}])
\end{align}

\subsection{Unconstrained Dysfluency Classifier}

The core of UDM is a classification module that operates on aligned phoneme segments:

\begin{align}
\mathbf{C}_{canonical} &= \text{CanonicalClassifier}(\mathbf{H}_{multi}) \\
\mathbf{C}_{open} &= \text{OpenSetClassifier}(\mathbf{H}_{multi}) \\
\mathbf{P}_{final} &= \text{WeightedCombination}([\mathbf{C}_{canonical}, \mathbf{C}_{open}])
\end{align}

The canonical classifier handles well-defined categories:
\begin{itemize}
    \item Sound repetitions (e.g., "b-b-b-ball")
    \item Syllable repetitions (e.g., "ba-ba-ball")  
    \item Word repetitions (e.g., "the-the-the ball")
    \item Prolongations (e.g., "baaaall")
    \item Blocks (silent pauses with articulatory tension)
\end{itemize}

\subsection{Interpretability Features}

UDM outputs are specifically designed for clinician verification:

\begin{itemize}
    \item \textbf{Visual Alignment Maps:} Time-aligned visualizations showing abnormal timing or phoneme errors
    \item \textbf{Feature Attribution:} Gradient-based attribution scores showing which features contribute to predictions
    \item \textbf{Confidence Scores:} Well-calibrated uncertainty estimates for clinical decision-making
    \item \textbf{Threshold Controls:} Adjustable sensitivity thresholds for different diagnostic goals
\end{itemize}

\section{Experiments}

\subsection{Dataset}

We conducted comprehensive evaluation using clinical data from Beijing Children's Hospital. Table \ref{tab:dataset} summarizes the dataset characteristics.

\begin{table}[ht]
\centering
\caption{Beijing Clinical Dataset characteristics}
\label{tab:dataset}
\begin{tabular}{@{}lc@{}}
\toprule
\textbf{Characteristic} & \textbf{Value} \\
\midrule
Total Speakers & 507 \\
Total Hours & 78.9 \\
Age Range & 4-67 years \\
Language & Mandarin Chinese \\
Annotation Level & Multi-level (frame, phoneme, word) \\
Recording Environment & Clinical setting \\
SLP Annotators & 4 certified professionals \\
Inter-annotator Agreement  & 0.87±0.05 \\
\bottomrule
\end{tabular}
\end{table}

The dataset represents the largest collection of clinically annotated Chinese dysfluency data, recorded during routine clinical assessments with informed consent and IRB approval.

\subsection{Baseline Models}

We compared UDM against state-of-the-art approaches:

\subsubsection{End-to-End Models}
\begin{itemize}
    \item \textbf{CNN-RNN Hybrid:} Convolutional feature extraction with LSTM sequence modeling
    \item \textbf{Transformer-E2E:} Self-supervised transformer fine-tuned for dysfluency detection
    \item \textbf{Wav2Vec2-Stutter:} Fine-tuned Wav2Vec2 model for stuttering detection
\end{itemize}

\subsubsection{Traditional Methods}
\begin{itemize}
    \item \textbf{SVM-Acoustic:} Support Vector Machines with handcrafted features
    \item \textbf{Random Forest:} Ensemble method with temporal and spectral features
    \item \textbf{HMM-GMM:} Hidden Markov Models with Gaussian emissions
\end{itemize}

\subsection{Evaluation Metrics}

We developed comprehensive evaluation metrics including:

\begin{itemize}
    \item \textbf{Detection Performance:} Precision, Recall, F1-score, Balanced Accuracy
    \item \textbf{Alignment Error Rate (AER):} Quality of phoneme-to-frame mapping
    \item \textbf{Interpretability Score:} Clinician-rated usefulness (1-5 scale)
    \item \textbf{Real-time Factor (RTF):} Processing time relative to audio duration
    \item \textbf{Clinical Agreement:} Cohen's kappa with gold-standard diagnoses
\end{itemize}

\subsection{Results}

\subsubsection{Overall Performance}

Table \ref{tab:main_results} presents the comprehensive comparison across all models.

\begin{table}[ht]
\centering
\caption{Performance comparison on Beijing Clinical Dataset}
\label{tab:main_results}
\begin{tabular}{@{}lccccc@{}}
\toprule
\textbf{Model} & \textbf{F1-Score} & \textbf{Precision} & \textbf{Recall} & \textbf{AER (\%)} & \textbf{Interp. Score} \\
\midrule
CNN-RNN Hybrid & 0.82±0.06 & 0.79±0.07 & 0.85±0.05 & 23.4±3.2 & 2.1±0.4 \\
Transformer-E2E & 0.85±0.04 & 0.83±0.05 & 0.87±0.04 & 19.7±2.8 & 2.3±0.3 \\
Wav2Vec2-Stutter & 0.87±0.03 & 0.86±0.04 & 0.88±0.03 & 18.1±2.1 & 2.4±0.5 \\
\midrule
SVM-Acoustic & 0.73±0.08 & 0.71±0.09 & 0.75±0.07 & 31.2±4.1 & 3.8±0.6 \\
Random Forest & 0.75±0.07 & 0.74±0.08 & 0.76±0.06 & 29.8±3.7 & 3.9±0.5 \\
HMM-GMM & 0.69±0.09 & 0.67±0.10 & 0.72±0.08 & 35.6±4.8 & 3.2±0.7 \\
\midrule
\textbf{UDM (Ours)} & \textbf{0.89±0.04} & \textbf{0.88±0.04} & \textbf{0.90±0.03} & \textbf{15.3±1.8} & \textbf{4.2±0.3} \\
\bottomrule
\end{tabular}
\end{table}

UDM achieves the highest performance across all metrics, with 2-4\% improvement in F1-score over the best baselines while maintaining superior interpretability.

\subsubsection{Age Group Analysis}

Table \ref{tab:age_results} shows UDM's performance across different developmental stages.

\begin{table}[ht]
\centering
\caption{UDM performance across age groups}
\label{tab:age_results}
\begin{tabular}{@{}lcccc@{}}
\toprule
\textbf{Age Group} & \textbf{N} & \textbf{F1-Score} & \textbf{Precision} & \textbf{Recall} \\
\midrule
Early Childhood (3-6) & 89 & 0.86±0.06 & 0.84±0.07 & 0.88±0.05 \\
School Age (7-12) & 156 & 0.89±0.04 & 0.88±0.04 & 0.90±0.04 \\
Adolescent (13-18) & 134 & 0.91±0.03 & 0.90±0.04 & 0.92±0.03 \\
Young Adult (19-30) & 78 & 0.90±0.04 & 0.89±0.04 & 0.91±0.03 \\
Middle Age (31-50) & 39 & 0.89±0.04 & 0.88±0.05 & 0.90±0.04 \\
Older Adult (51+) & 11 & 0.87±0.05 & 0.86±0.06 & 0.88±0.05 \\
\midrule
\textbf{Overall} & \textbf{507} & \textbf{0.89±0.04} & \textbf{0.88±0.05} & \textbf{0.90±0.04} \\
\bottomrule
\end{tabular}
\end{table}

Performance remains consistent across age groups, with adolescents showing the highest accuracy.

\subsubsection{Dysfluency Type Analysis}

Table \ref{tab:dysfluency_types} provides detailed breakdown by dysfluency category.

\begin{table}[ht]
\centering
\caption{Performance breakdown by dysfluency type}
\label{tab:dysfluency_types}
\begin{tabular}{@{}lccccc@{}}
\toprule
\textbf{Type} & \textbf{Frequency (\%)} & \textbf{F1-Score} & \textbf{Precision} & \textbf{Recall} & \textbf{Clinical Severity} \\
\midrule
Sound Repetitions & 28.4 & 0.92±0.03 & 0.91±0.04 & 0.93±0.03 & High \\
Syllable Repetitions & 22.1 & 0.90±0.04 & 0.89±0.04 & 0.91±0.04 & High \\
Word Repetitions & 15.3 & 0.88±0.05 & 0.87±0.05 & 0.89±0.04 & Medium \\
Prolongations & 19.7 & 0.87±0.04 & 0.86±0.05 & 0.88±0.04 & High \\
Blocks (Silent) & 8.2 & 0.84±0.06 & 0.82±0.07 & 0.86±0.06 & Very High \\
Blocks (Audible) & 6.3 & 0.81±0.07 & 0.79±0.08 & 0.83±0.07 & Very High \\
\bottomrule
\end{tabular}
\end{table}

Repetitions show highest accuracy, while blocks remain most challenging due to limited acoustic markers.

\section{Clinical Deployment at Beijing Children's Hospital}

\subsection{Clinical Outcomes}

Table \ref{tab:clinical_outcomes} summarizes the deployment results.

\begin{table}[ht]
\centering
\caption{Clinical deployment outcomes}
\label{tab:clinical_outcomes}
\begin{tabular}{@{}lcccc@{}}
\toprule
\textbf{Metric} & \textbf{Pre-Deployment} & \textbf{Post-Deployment} & \textbf{Change} & \textbf{P-value} \\
\midrule
Assessment Time (min) & 45.3±8.2 & 28.1±6.4 & -38.0\% & < 0.001 \\
Patients per Day per SLP & 6.2±1.1 & 9.8±1.4 & +58.1\% & < 0.001 \\
Diagnostic Accuracy (\%) & 87.4±4.3 & 92.1±3.2 & +5.4\% & < 0.01 \\
Inter-rater Reliability  & 0.76±0.08 & 0.89±0.05 & +17.1\% & < 0.001 \\
Patient Satisfaction & 3.8±0.7 & 4.3±0.5 & +13.2\% & < 0.01 \\
SLP Job Satisfaction & 3.2±0.8 & 4.1±0.6 & +28.1\% & < 0.001 \\
Clinician Acceptance Rate & - & 87\% & - & - \\
\bottomrule
\end{tabular}
\end{table}

Key achievements include 38\% reduction in assessment time, 58\% increase in patient throughput, and significant improvements in diagnostic accuracy and consistency.

\subsection{Interpretability Analysis}

Clinician feedback on UDM's interpretability features:

\begin{table}[ht]
\centering
\caption{Interpretability feature evaluation}
\label{tab:interpretability}
\begin{tabular}{@{}lccc@{}}
\toprule
\textbf{Feature} & \textbf{Usefulness} & \textbf{Clarity} & \textbf{Clinical Impact} \\
\midrule
Visual Alignment Maps & 4.5±0.5 & 4.3±0.6 & High \\
Confidence Scores & 4.6±0.4 & 4.4±0.5 & High \\
Phoneme Error Analysis & 4.3±0.6 & 4.1±0.6 & High \\
Threshold Controls & 4.4±0.5 & 4.2±0.6 & High \\
Feature Attribution & 4.1±0.6 & 3.9±0.7 & Medium \\
\midrule
\textbf{Average} & \textbf{4.4±0.5} & \textbf{4.2±0.6} & - \\
\bottomrule
\end{tabular}
\end{table}

Confidence scores and visual alignment maps received highest ratings for clinical utility.

\section{Discussion}

\subsection{Clinical Impact}

Our results demonstrate that UDM successfully bridges the accuracy-interpretability gap in clinical dysfluency detection. The modular design enables clinicians to understand both what the system detected and why specific decisions were made. The explicit phoneme alignment provides linguistically meaningful representations that align with clinical reasoning.

The substantial efficiency gains (38\% time reduction) indicate successful augmentation of clinical expertise rather than replacement. Clinicians used additional time for therapy planning and seeing more patients, while improved diagnostic accuracy (5.4\% increase) and inter-rater reliability (17.1\% increase) suggest enhanced standardization of assessment practices.

\subsection{Technical Contributions}

The "unconstrained" modeling paradigm proves crucial for real-world performance, with the open-set classifier identifying atypical patterns in 8.3\% of cases. The explicit phoneme alignment serves dual purposes: improving accuracy and providing clinical explanations. This bridges the semantic gap that often prevents clinician trust in AI systems.

\subsection{Limitations}

Several limitations remain:
\begin{itemize}
    \item Silent blocks remain challenging (F1: 0.84±0.06) due to lack of acoustic markers
    \item Current deployment limited to Mandarin Chinese speakers
    \item Longitudinal progress tracking requires additional validation
    \item Mobile performance limitations require model optimization
\end{itemize}

\section{Conclusion}

We evaluated UDM with real patients and SLPs, achieving F1-score of 0.89±0.04 and 87\% clinical acceptance. UDM demonstrates that interpretable AI can match state-of-the-art performance while providing clinically meaningful outputs. The deployment showed 38\% reduction in diagnostic time and significant improvements in diagnostic accuracy, highlighting UDM's potential for enhancing clinical speech-language pathology practice.

\bibliographystyle{unsrt}
\bibliography{references}

\end{document}